\documentclass[pre,twocolumn,showpacs,floatfix]{revtex4}
\usepackage{graphicx,amssymb,amsmath,mathptm}
\begin{document}
\title{Stationary and dynamical properties of a zero range process on scale-free networks}
\author{Jae Dong Noh}
\affiliation{Department of Physics, Chungnam National University, Daejeon
305-764, Korea}
\date{\today}

\begin{abstract}
We study the condensation phenomenon in a zero range process on 
scale-free networks. We show that the stationary state property depends
only on the degree distribution of underlying networks.
The model displays a stationary state phase transition
between a condensed phase and an uncondensed phase, and the phase diagram is
obtained analytically.
As for the dynamical property, we find that the relaxation dynamics depends
on the global structure of underlying networks.
The relaxation time follows the power law $\tau \sim L^z$ with the network 
size $L$ in the condensed phase.
The dynamic exponent $z$ is found to take a different value depending on 
whether underlying networks have a tree structure or not.

\end{abstract}

\maketitle

\section{Introduction}\label{sec:intro}
A condensation is an intriguing phenomenon observed in nonequilibrium
systems~\cite{Evans00}. In interacting particle systems, a finite
fraction of particles may be condensed onto a single site.
Phase separation in driven diffusive systems~\cite{Evans98,Arndt98,Kafri02}
or jamming in traffic flows~\cite{Evans00} is also an
example of the condensation.
The condensation has been studied within the framework of a zero range
process~(ZRP) introduced by Spitzer~\cite{Spitzer70}.
Those studies revealed that the condensation can be caused by
effective attractions between particles or quenched 
disorder~\cite{Evans00,Grosskinsky03,Hanney05,Krug96,Evans96,Krug00,Jain03}. 
In the former the condensation is a spontaneous symmetry breaking transition. 
In the latter it is similar to the Bose-Einstein condensation
occurring in the ideal Bose gas~\cite{Hwang}.

The condensation has been studied mostly on periodic lattices.
On the other hand, recent studies show that
lattices~(networks in general), upon which physical systems are defined,
may have more complex structure~\cite{Watts98,Albert02,Dorogov02,Newman03}.
Particularly many real-world networks are found to be {\em scale-free}.
That is to say, they are characterized with the power-law degree distribution
\begin{equation}\label{P_deg}
P_{\text{deg}}(k) \sim k^{-\gamma} \ ,
\end{equation}
where $P_{\text{deg}}(k)$ is the fraction of nodes~(or sites) with the degree
$k$~(the degree of a node denotes the number of neighbors connected to
it) and $\gamma$ is called the degree distribution exponent.
This is contrasted to a peaked distribution for periodic regular lattices.

The power-law distribution implies that the SF networks have an inhomogeneous
structure.
The structural inhomogeneity is an important factor in understanding
dynamical processes in SF
networks~\cite{Barthelemy04,Noh04a,Noh04b,Gallos04,Catanzaro05,Sood05}.
In random walks, for instance, a diffusing particle
visits highly connected nodes more often than others.
Moreover diffusion processes are asymmetric.
It means that the speed of random walks from a node to another is not equal
to that in the reversed direction in general.
The asymmetry can be quantified with the so-called random walk
centrality~\cite{Noh04a}, which is defined on each node and proportional to
its degree.
This example shows that each node plays a different role in stochastic
processes in inhomogeneous networks.
It also suggests that interactions in many particle systems would be
important since particle density could be high at some nodes.

In this paper, we study a ZRP on SF networks and investigate the
interplay of the dynamics of interacting particle systems and the
inhomogeneous structure of underlying networks. The present work is an
extension of our recent work presented in Ref.~\cite{Noh05}, which shows
that the structural inhomogeneity gives rise to the novel type of
condensation in the ZRP.
We will present a detailed review on the results. In addition, we will
elaborate a scaling theory for the relaxation dynamics in the condensed
phase. Main results are following.
The relaxation dynamics has a {\em hierarchical} nature; 
starting from small degree
nodes, the relaxation proceeds to larger and larger degree nodes.
The relaxation time $\tau$ follows a power-law scaling
$\tau \sim L^z$ with the dynamic exponent $z$ and the network size $L$
in the condensed phase. The relaxation dynamics depends 
on the global structure of underlying networks so that
the dynamic exponent takes a different value depending on whether the
networks have a tree structure or not. Usually, the degree distribution
is considered to be important in characterizing the dynamics of complex 
networks. Our result shows that the global structure of networks is an
important ingredient in understanding the dynamics.

This paper is organized as follows. In Sec.~\ref{sec:2}, we provide a
general review on stationary and dynamical properties of the ZRP.
Then in Sec.~\ref{sec:3} we introduce the ZRP on SF networks and
present its stationary state phase diagram.
In Sec.~\ref{sec:4} we investigate the relaxation dynamics.
We summarize and conclude the paper in Sec.~\ref{sec:5}.

\section{Zero Range Process}\label{sec:2}
Consider a system of $N$ interacting particles on a network of $L$ nodes.
The total particle density is given by $\rho = N/L$.
Each node $i$ can be occupied by any integral number of particles~(no
exclusion). With $n_i=0,1,2,\cdots$ denoting the occupation number
at node $i$, a microscopic configuration is represented with
${\mathbf n} = (n_1,n_2,\cdots ,n_L)$.

A ZRP is specified by jumping rates $\{p_i(n)\}$ and hopping probabilities
$\{T_{j\leftarrow i}\}$. With those, at each node $i$ with $n_i>0$, a
particle jumps out at the rate $p_i(n_i)$ and then hops to a node $j$
with the probability $T_{j \leftarrow i}$.
Note that the jumping rate depends only on the occupation number at 
departing nodes. This is an important feature of the ZRP, which makes
the exact stationary state probability distribution available~\cite{comment3}.
Recently, generalizations with multi-species particles~\cite{Evans03}
or with multiple-particle hoppings~\cite{Evans04}
are being considered. 

A particle interaction can be incorporated into the jumping rate. If
particles are noninteracting, they move independently, hence $p(n) \propto
n^1$. An attractive interaction can be imposed with the jumping
rate function $p(n)$ that grows sub-linearly or decreases with $n$,
or a repulsive interaction with a super-linear jumping rate function.
In the context of data packet transport, the jumping rate $p_i(n)$ corresponds
to the number of packets that can be processed by a node $i$ having $n$ 
packets per unit time. 
That is, the jumping rate function models the transport capacity.

It is well known that the stationary state property of the ZRP is exactly
solvable~\cite{Evans00,Hanney05}. For later
uses, we list useful formula below.
The stationary
state probability distribution $P_{\text{stat}}({\mathbf n})$ is
given by
\begin{equation}\label{P_n}
P_{\text{stat}}({\mathbf n}) = \frac{1}{Z(L,N)} \prod_{i=1}^L f_i(n_i) \ ,
\end{equation}
where
\begin{equation}\label{f_n}
f_i(n) = \prod_{n'=1}^n \left(\frac{\omega_i}{p_i(n')} \right)
\end{equation}
for $n\neq0$ and $f_i(0)=1$, and
$Z(L,N)$ is a normalization constant. 
Here, $\{\omega_i\}$ is the solution of coupled linear equations
\begin{equation}\label{omega}
\sum_j T_{j\leftarrow i}\ \omega_i = \sum_{j}\ T_{i\leftarrow j}\ \omega_j
\quad (i=1,\cdots,L) .
\end{equation}
Note that these equations are equivalent to those for the stationary state
probability distribution for a {\em single} random walker problem with the
hopping probability $\{T_{j\leftarrow i}\}$. The left~(right) hand side
corresponds to the outgoing~(incoming) probability flux.

Introducing an auxiliary fugacity variable $z$~\cite{Evans00},
one finds that the mean occupation number $m_i = \langle n_i \rangle$
is given by
\begin{equation}\label{m_i_z}
m_i = z \frac{\partial}{\partial z} \ln F_i(z) \ ,
\end{equation}
where
\begin{equation}\label{F_i_z}
F_i(z) \equiv \sum_{n=0}^{\infty} z^n f_i (n) \ .
\end{equation}
The series is defined for $|z|<z_c$ with $z_c$ being 
the radius of convergence.
The fugacity should be determined from the self-consistent equation
\begin{equation}\label{self_consistent_eq}
\rho = \frac{1}{L} \sum_{i=1}^L m_i(z) \ .
\end{equation}

A ZRP may display the condensation transition. We introduce two instances.
A simplest ZRP displaying the condensation might be the one
with the jumping rate
$ p_i(n) = (1-\delta_{i,i_0}) + \alpha \delta_{i,i_0} $
on a periodic regular lattice with nearest neighbor hoppings~\cite{Evans00}.
This model has a quenched disorder so that
the impurity node $i_0$ has a low jumping rate $\alpha<1$.
Applying the formalism, one can show easily that the condensation occurs for
$\rho > \rho_c = \alpha/(1-\alpha)$, where a macroscopic 
condensate forms at the impurity site with the occupation number 
$m_{i_0} = (\rho-\rho_c)L$, while the occupation number at the normal nodes
is given by $m_{i\neq i_0}= \rho_c$.

The condensation also occurs in a system without quenched disorder.
For instance, consider a ZRP on periodic regular lattices
with the jumping rate $p_i(n) = p(n) = 1+b/n$.
A straightforward analysis shows that the system is in the condensed
phase when $b>2$ and $\rho>\rho_c=\rho_c(b)$~\cite{Evans00,Grosskinsky03}.
In this case, a macroscopic condensate forms at a node selected 
spontaneously.

As for the relaxation dynamics, an exact theory has not been known yet.
However, qualitative features may be understood from a scaling 
argument~\cite{Grosskinsky03,Godreche03,Torok04,Grosskinsky04}. To be
specific, we consider the relaxation dynamic of the above example 
with the jumping rate $p(n) = 1+b/n$ in the condensed phase. 
Being distributed uniformly initially, particles form small clusters at
nodes due to a stochastic fluctuation. It is believed that a macroscopic
condensate emerges by successive coarsening processes of the small clusters.
Those clusters exchange particles, merge into larger ones, and grow until 
there remains a single macroscopic condensate. 
This scaling picture predicts the power-law growth of the cluster
size in time and the power-law scaling of the relaxation time with the
system size as $\tau \sim L^z$. It is known that the dynamic exponent $z$ 
depends on the dimensionality of the underlying lattices and bias in 
particle hoppings~\cite{Grosskinsky03,Godreche03}. 

\section{ZRP on SF networks}\label{sec:3}
A recent study reveals that structural inhomogeneity of 
SF networks leads to the so-called {\em complete
condensation}~\cite{Noh05}. Being compared to the usual condensation
in periodic regular lattices~\cite{Evans00}, the complete condensation
has the features that (i) it occurs at any finite
value of the particle density, i.e., $\rho_c =0$, and that
(ii) the whole fraction of particles are condensed onto a few high-degree
nodes. We review these results in this section.

Consider a ZRP on a SF network of $L$ nodes. The network is assumed to have
a power-law degree
distribution $P_{\text{deg}}(k)\sim k^{-\gamma}$ in the interval $k_0
\le k \le k_{\max}$. We are interested in sparse networks with finite mean
degree $\bar{k} \equiv \int dk \; k P_{\text{deg}}(k)$. Hence, the degree
distribution exponent is assumed to be larger than $2$, $\gamma>2$.

The node with the maximum degree $k_{\max}$ is 
called the {\em hub}. While the lower bound $k_0$ is a constant,
the maximum degree $k_{\max}$ scales with the network size $L$.
It can be estimated from the condition
$L \int_{k_{\max}}^\infty P_{\text{deg}}(k) dk = 1$,
which yields that~\cite{Dorogov02}
\begin{equation}\label{beta}
k_{\max} \sim L^{\beta} \mbox{\ \ with \ \ } \beta=\frac{1}{\gamma-1} \ .
\end{equation}
Some SF networks having an explicit cutoff for $k_{\max}$ may have a
different value of $\beta$~\cite{cutoff}. In this work, however,
we only consider the generic SF networks with $\beta$ given by
Eq.~(\ref{beta}).

For the jumping rate, we take
\begin{equation}
p_i(n) = n^\delta
\end{equation}
for all nodes with a parameter $\delta \geq 0$. We take
\begin{equation}\label{T_ij_net}
T_{j\leftarrow i}  = \left\{
  \begin{array}{lll}
 \displaystyle{\frac{1}{k_i}} & , & \mbox{if $i$ and $j$ are linked} \\
  0             &,& \mbox{otherwise}
  \end{array}
\right.
\end{equation}
for the hopping probability~($k_i$ is the degree of a node $i$).
That is to say, a particle out of $i$ hops to one of its $k_i$ neighbors
selected at random.

First, one needs to solve the random walk problem in Eq.~(\ref{omega}) for
the stationary state probability $P_{\text{stat}}(\mathbf{n})$.
The random walk problem with the hopping probability (\ref{T_ij_net})
was studied in Ref.~\cite{Noh04a}, from which we find
that~\cite{comment1}
\begin{equation}\label{omega_k}
\omega_i = k_i \ .
\end{equation}
Inserting $p(n)$ and $\{\omega_i\}$ into Eq.~(\ref{f_n}), one finds
that
\begin{equation}\label{eq:fin}
f_i(n) = \prod_{l=1}^n \ \left( \frac{k_i}{l^\delta} \right) =
\frac{k_i^n}{(n!)^\delta} \ ,
\end{equation}
which further yields from Eq.~(\ref{F_i_z}) that
\begin{equation}
F_i(z) = \sum_{n=0}^\infty \frac{ (z k_i )^n }{ (n!)^\delta } \ .
\end{equation}
Hence, the mean occupation number in the stationary state is given from
Eq.~(\ref{m_i_z}) by
\begin{equation}\label{m_z_net_formal}
m_i(z) = x \left. \frac{\partial \ln \mathcal{F}_\delta (x)}{\partial x}
\right|_{x=zk_i} \ ,
\end{equation}
where
\begin{equation}\label{F_x}
{\mathcal F}_\delta(x) \equiv \sum_{n=0}^\infty \frac{x^n}{(n!)^\delta} \ .
\end{equation}
The fugacity $z$ has to be determined from the self-consistent equation.

We add a few remarks on the stationary state probability distribution.
Although the jumping rate is the same at all nodes, the probability
factor $f_i(n)$ in Eq.~(\ref{eq:fin}) has the node dependence due to the
structural inhomogeneity of underlying networks. 
On the other hand, consider a ZRP in {\em periodic lattices} 
with {\em disordered} jumping rates $p_i'(n) = n^\delta / k_i$ and 
with nearest neighbor particle hoppings.
It is evident that the two models share the same stationary state.
That is, the {\em structure inhomogeneity} plays the same role as the 
{\em quenched disorder} in the jumping rate.
At $\delta=0$, in particular, the corresponding ZRP in periodic
lattices is characterized by the disordered jumping rate $p'=1/k$
whose distribution is given by 
\begin{equation}
P_{\text{jump}}(p') =
P_{\text{deg}}(k)\ \left| \frac{dk}{dp'} \right| \sim p'^{(\gamma-2)} \ .
\end{equation}
Such a problem has been studied in the context of a
disordered ZRP and a disordered exclusion process in one
dimension~\cite{Krug96,Evans96,Krug00,Jain03}. In those studies, the jumping
rate distribution is taken to be $P_{\text{jump}}(c<p'<1)\sim (p'-c)^\nu$ with 
the model parameter $\nu$ and $c$ for the minimum jumping rate. 
The present model at $\delta=0$ coincides with that with $c=0$ and $\nu =
\gamma-2$.

It is noteworthy that the mean occupation number in the stationary state
is solely determined by the degree, which is a local quantity, and
independent of any other characteristics of a network. 
This, however, is not the case for the relaxation dynamics studied 
in Sec.~\ref{sec:4}.

At $\delta=1$, the particle jumping rate is directly proportional to the
occupation number. Particles move independently and 
the system reduces to a non-interacting system of $N$ random walkers.
So, the stationary state occupation number distribution becomes 
proportional to the single walker distribution in Eq.~(\ref{omega_k}), 
$m_i \propto \omega_i =k_i$.
This solution is also obtained from the general formula in
Eq.~(\ref{m_z_net_formal}). Since $\mathcal{F}_1(x) = e^x$, $m_i(z) = zk_i$,
which is proportional to the degree as expected.

For general values of $\delta$, the solution is non-trivial.
We will solve the problem for $\delta=0$ and $\delta\neq0$ separately.

\subsection{$\delta=0$ case}

At $\delta=0$, the infinite series for $\mathcal{F}_\delta(x)$ has a closed
form expression $\mathcal{F}_0 (x) = 1 / (1-x)$, which yields
\begin{equation}
m_i(z) = \frac{z k_i }{1-z k_i } \ .
\end{equation}
The fugacity is in the range $z<z_c = 1 / k_{\max}\ll 1$.

Note that $m_i(z)$ is divergent for the hub as $z$ approaches $z_c$.
So, it is convenient to decompose the self-consistent equation as
$\rho = \rho_s + \rho_n$ where
\begin{equation}
\rho_s = \frac{m_{\text{hub}}}{L} = \frac{1}{L}\frac{z k_{\max}}{1-zk_{\max}}
\end{equation}
is the particle density at the hub and
\begin{equation}
\rho_n = \frac{1}{L} \sum_{i\neq \text{hub}} \frac{z k_i }{1-z k_i }
\end{equation}
is the remaining density.
For large $L$, one can evaluate $\rho_n$ as
$$ \rho_n = \int_{k_0}^{k_{\max}} dk\; \frac{P_{\text{deg}}(k) zk}{1-zk}
= z^{\gamma-1} \int_{k_0 z}^{k_{\max} z} dx \;\frac{x^{1-\gamma}}{(1-x)} \ .$$
The integrand becomes singular at both integral endpoints
as $zk_{\max}\rightarrow 1$ and $zk_0 \rightarrow 0$.
Evaluating the singular parts, we obtain that
$\rho_n = \mathcal{O}(z) + \mathcal{O}( z^{\gamma-1} \ln (1-z
k_{\max}))$, where the first~(second) term is a contribution
near the lower~(upper) endpoint~\cite{comment2}.
Since $z < 1/k_{\max}\sim L^{-\beta} \ll 1$ and $\gamma>2$, 
$\rho_n$ vanishes in the large $L$ limit. 
Therefore we conclude that the whole fraction of particle should
be condensed into the hub~[complete condensation],
\begin{equation}
\rho_s = m_{\text{hub}} / L = \rho \ .
\end{equation}
The fugacity is then given by
\begin{equation}
z = k_{\max}^{-1} / ( 1 + 1/(\rho L)) \sim L^{-\beta} ( 1 - 1/(\rho L))
\ .
\end{equation}
So the mean occupation number at other nodes with $k_i \ll k_{\max}$ is
given as
\begin{equation}
m_{i\neq \text{hub}} \simeq zk_i \simeq k_i / k_{\max} \sim L^{-\beta} k_i \ ,
\end{equation}
which is proportional to the degree with the vanishing coefficient.

\subsection{$\delta>0$ case}
For $\delta>0$, the infinite series in Eq.~(\ref{F_x})
does not have a closed form expression except for
$\mathcal{F}_{\delta=1}(x) = e^x$.
We approximate the series in the following way.
Putting the summand as $x^n/(n!)^\delta\equiv \exp[g(n)]$ and
using the Stirling formula, one obtains that
$$ g(n) \simeq n \ln x - \delta \left\{ \frac{1}{2}\ln(2\pi n) + n \ln n -
n\right\} \ . $$
This function is peaked at $n=n_0$ and behaves near $n\simeq n_0$ as
$$ g(n) \simeq g(n_0) + \frac{1}{2} g''(n_0) (n-n_0)^2 + \cdots \ .$$
The peak position is easily found to be $n_0 \simeq x^{1/\delta}$ from the
condition $g'(n_0)=0$, which yields
$g''(n_0) = -\delta/n_0 + \mathcal{O}(n_0^{-2})$.
We then approximate the series with the Gaussian integral as
\begin{eqnarray}\label{F_app}
\mathcal{F}_{\delta> 0}(x) &\simeq& e^{g(n_0)} \int_{-n_0}^\infty dn'
\exp\left[ g''(n_0) n'^2 /2 \right]  \nonumber \\
&\simeq& e^{g(n_0)} \int_{-\infty}^{\infty} dn'
 \exp\left[ g''(n_0) n'^2 \right] \nonumber \\
&\simeq& \frac{1}{\sqrt{\delta}} \left( 2\pi x^{1/\delta}
\right)^{(1-\delta)/2} \exp\left( \delta x^{1/\delta}\right)  \ .
\end{eqnarray}
In the second step, we extend the integration interval.
This approximation is valid when
$ |g''(n_0)| n_0^2 = \delta x^{1/\delta} \gtrsim 1$,
that is $x \gtrsim (1/\delta)^\delta=\mathcal{O}(1)$.
For $x\ll 1$, one can simply approximate
$\mathcal{F}_\delta(x)$ with a few lowest order terms as
$\mathcal{F}_\delta(x) = 1+x +\mathcal{O}(x^2)$.

\begin{figure}[t]
\includegraphics*[width=\columnwidth]{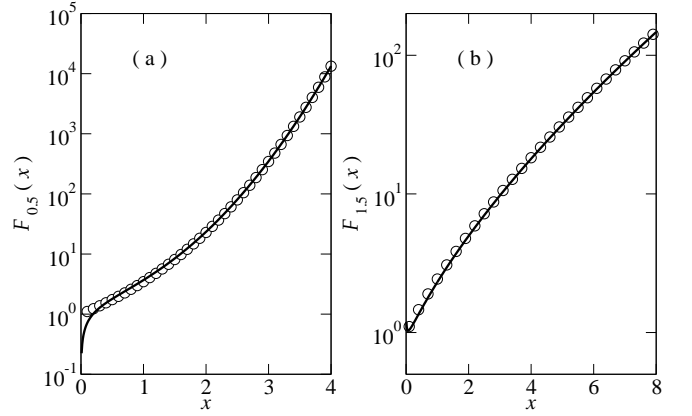}
\caption{Comparison of $\mathcal{F}_\delta(x)$~(circular symbols) defined
in Eq.~(\ref{F_x}) and the approximation~(solid line)
in Eq.~(\ref{F_app}) at $\delta=0.5$~(a) and 1.5~(b).}
\label{fig1}
\end{figure}

Note that the approximation in Eq.~(\ref{F_app}) coincides with
the exact result at $\delta=1$. We have tested its validity
at other values of $\delta$ numerically. Figure~\ref{fig1} shows
that the approximation is very good unless $x\simeq 0$.

Using the approximation for $\mathcal{F}_{\delta}(x)$,
the mean occupation number is given by
\begin{equation}\label{m_i_net_cond}
m_i \simeq \left\{
 \begin{array}{ll}
   z k_i  & \mbox{\quad for\quad } z k_i \ll 1 \\ [1mm]
   \displaystyle (z k_i )^{1/\delta} & \mbox{\quad for\quad } z k_i \gtrsim 1
 \ .
 \end{array}
\right.
\end{equation}
The mean occupation number of a node increases monotonically with its degree.
So we expect that $m_{\text{hub}} \gtrsim 1$ and that
$z k_{\max} \gtrsim 1$. On the other hand, for a
node with the minimum degree $k_0$, one may have $z k_0 \ll 1$ or $z k_0 
\gtrsim 1$ depending on the magnitude of $z$. Hence, one must consider
the following two cases separately.

(i) Firstly, let us assume that the fugacity is in such a range that
$z k_i \gtrsim 1 $ for all nodes. This is the case when $z$ is a
non-vanishing constant in the large $L$ limit.
Then the occupation number is given by $m_i \simeq (k_iz)^{1/\delta}$ for
all nodes, and the self-consistent equation has the solution
$z= \rho^\delta / \left(\overline{k^{1/\delta}}\right)^\delta$
where $\overline{k^{1/\delta}} \equiv \int_{k_0}^{k_{\max}} dk \;
k^{1/\delta} P_{\text{deg}}(k)$.
It could be a non-vanishing constant only when
$\overline{k^{1/\delta}} \sim \int_{k_0}^{k_{\max}} dk k^{-(\gamma-1/\delta)}$
remains finite in the large $L$ limit. It imposes that $\delta >
\delta_c$ with
\begin{equation}\label{delta_c}
\delta_c = \frac{1}{\gamma-1} \ .
\end{equation}
In this regime, we find that
\begin{equation}\label{m_i_noncond}
m_i \simeq c k_i^{1/\delta}
\end{equation}
for all nodes with a constant $c=\rho /\left({\overline{k^{1/\delta}}}\right)$.
The occupation number at the hub with the degree $k_{\max}\sim L^{\beta}$
scales as $m_{\text{hub}}\sim L^{\beta/\delta}$. Since
$\beta=1/(\gamma-1)$ and $\delta>\delta_c = 1/(\gamma-1)$, the occupation number at the hub scales
sub-linearly in $L$. Therefore, the condensation does not occur when
$\delta>\delta_c$.

(ii) Secondly, let us assume that the fugacity $z$ is in such a range that
the {\em crossover degree} defined as $k_c \equiv 1/z$ scales with the
network size $L$ and is in the interval $k_0 \ll k_c \ll k_{\max}$.
Then, the self-consistent equation becomes $\rho = \rho_n + \rho_s$ where
\begin{equation}\label{rho_n_net}
\rho_n = k_c^{-1} \; \int_{k_0}^{k_c} dk \; k P_{\text{deg}}(k)
\end{equation}
is the density at small-degree nodes with $k < k_c$ and
\begin{equation}\label{rho_s_net}
\rho_s = k_c^{-1/\delta} \; \int_{k_c}^{k_{\max}} dk \; 
k^{1/\delta} P_{\text{deg}}(k)
\end{equation}
is the density at large-degree nodes with  $k> k_c$.
The integral part in Eq.~(\ref{rho_n_net}) is smaller than the mean degree
$\bar{k}$ that is finite~$(\gamma>2)$. So, $\rho_n$ vanishes as 
$\sim k_c^{-1}$ in the large $L$ limit, which yields that $\rho_s = \rho$.
In order to have a finite value of $\rho_s$, 
$\int_{k_c}^{k_{\max}} dk k^{1/\delta-\gamma}$ should be
divergent, which yields that $\delta \le \delta_c =
1/(\gamma-1)$. And it should be of the same order as
$k_c^{-1/\delta}$, which yields
\begin{equation}\label{k_crossover}
k_c \sim \left\{
\begin{array}{lll}
 \left[ \ln k_{\max} \right]^{\delta_c} &,& \mbox{ for \quad} \delta =
\delta_c \\ [2mm]
 \left[ k_{\max} \right]^{1-\delta/\delta_c} &,& \mbox{ for \quad} \delta <
\delta_c \ .
\end{array} \right.
\end{equation}
With the crossover degree $k_c=1/z$, the occupation number distribution
is given by Eq.~(\ref{m_i_net_cond}),which can be summarized with the 
scaling form
\begin{equation}\label{m_cond_scale}
m_i = G_{\delta} ( k_i / k_c )
\end{equation}
with the scaling function $G_\delta(y)$ that behaves as
$G_{\delta}(y\ll 1) \sim y$ and
$G_{\delta}(y\gtrsim 1) \sim y^{1/\delta}$.

We add a few remarks on the result in Eq.~(\ref{m_cond_scale}).
Nodes with $k_i<k_c$ covers the whole fraction of nodes in a network since
$\int_{k_c}^{k_{\max}} dk P_{\text{deg}}(k) \sim k_c^{-(\gamma-1)}$
vanishes in the $L\rightarrow \infty$ limit. However, the total density of
particles on those nodes, given by $\rho_n\sim k_c^{-1}$,
vanishes in the thermodynamic limit.
Therefore, the whole fraction of particles are condensed onto the
vanishingly small fraction of nodes with $k_i >k_c$. For those nodes, the
occupation number is given by
$m_i \simeq (k_i/k_c)^{1/\delta} \sim (k_{\max}/k_c)^{1/\delta}
(k_i/k_{\max})^{1/\delta}$. It has the overall $L$ dependence as
$(k_{\max}/k_c)^{1/\delta} \sim L^{\beta/\delta_c} \sim L$
for $\delta<\delta_c$ and $\sim L/ \ln L$ for $\delta=\delta_c$.
That is, the occupation number scales linearly with the system size $L$ for
$\delta<\delta_c$. One can also find that the scaling form in
Eq.~(\ref{m_cond_scale}) can be applied to the case of $\delta=0$ studied in
the previous subsection with $k_c = k_{\max}$.
Therefore, we conclude that there is the complete
condensation for $\delta<\delta_c$. At the threshold $\delta = \delta_c$,
the logarithmic correction appears.

As explained earlier, the stationary state of the present model 
is equivalent to that of the disordered ZRP on periodic lattices 
with the jumping rate $p_i'(n) = n^\delta/k_i$. 
The disordered ZRP with $n$-independent jumping rates, which corresponds to
the $\delta=0$ case, has been studied in 
Refs.~\cite{Krug96,Evans96,Krug00,Jain03}. Those studies show that
the condensation is triggered by the node with the slowest
hopping rate and the complete condensation occurs when the slowest jumping
rate vanishes~\cite{Krug96,Evans96,Krug00,Jain03}. 
Generalizing the results, one can understand our phase diagram intuitively. 
The slowest hopping rate node corresponds to the hub with the maximum 
degree $k_{\max}\sim L^\beta$. 
Assuming that the node is occupied macroscopically, one finds that 
the jumping rate is given by $p' = n^\delta / k_i \sim L^{\delta - \beta}$. 
It vanishes when $\delta < \beta = 1/(\delta-1)$, which coincides with the
phase boundary.

\begin{figure}[t]
\includegraphics*[width=.9\columnwidth]{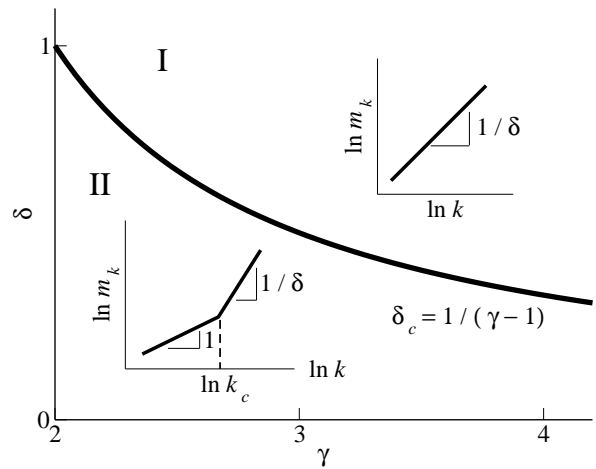}
\caption{Stationary state phase diagram. The region I and II corresponds to
the uncondensed and condensed phase, respectively.}\label{fig2}
\end{figure}

Summarizing the results, we present the stationary
state phase diagram of the ZRP on SF networks in
Fig.~\ref{fig2}. The region I corresponds to the uncondensed phase,
where the occupation number is given by
Eq.~(\ref{m_i_noncond}).
The region II corresponds to the condensed phase,
where the occupation number is given by
Eq.~(\ref{m_cond_scale}).
The phase boundary is given by $\delta=\delta_c$ with
Eq.~(\ref{delta_c}).
The phase diagram suggests that the
condensation also occurs in random networks~($\gamma=\infty$) at
$\delta=0$, which is observed in a numerical study~\cite{YKim05}. The
occupation number distribution in each phase is sketched in
Fig.~\ref{fig2}.

\section{Relaxation dynamics}\label{sec:4}
The preceding analysis shows that the occupation number distribution
in the stationary state is solely determined by the degree
distribution $P_{\text{deg}}(k)$ being independent of any other
characteristics of underlying networks. However, the relaxation
dynamics may be affected by the structure of underlying networks. 
We are interested in the relaxation dynamics in the condensation phase.
Unlike the stationary state property, an exact theory
is not available for the dynamical property of the ZRP. We rely on
Monte Carlo simulations for the study and provide a scaling argument
to understand dynamical scaling properties.

We describe our Monte Carlo simulation algorithm. Initially at time
$t=0$, $N$ particles are distributed uniformly and randomly on a
network of $L$ nodes. At each step, a {\em particle} is selected at
random. Let $i$ be its residing site and $n_i$ be the occupation
number there. Then, with the probability
$p(n_i)/n_i=n_i^{\delta-1}$, the particle jumps out of $i$ and hops
to one of its neighboring nodes. With the probability
$1-p(n_i)/n_i$, the selected particle does not move. The time is
measured in unit of one Monte Carlo sweep consisting of $N$ trials.
Note that the particle-based dynamics with the modified jumping
probability $p(n)/n$ is equivalent to the original node-based
dynamics with the jumping rate $p(n)$.

As for SF networks, we use the Dorogovtsev-Mendes-Samukhin
model~\cite{DMS} that is a generalization of the Barab\'asi-Albert
model~\cite{BAnet}. With the model, one can generate networks of tree
structure or looped structure with arbitrary values of 
the degree distribution exponent $\gamma$.

\begin{figure}[t]
\includegraphics*[width=\columnwidth]{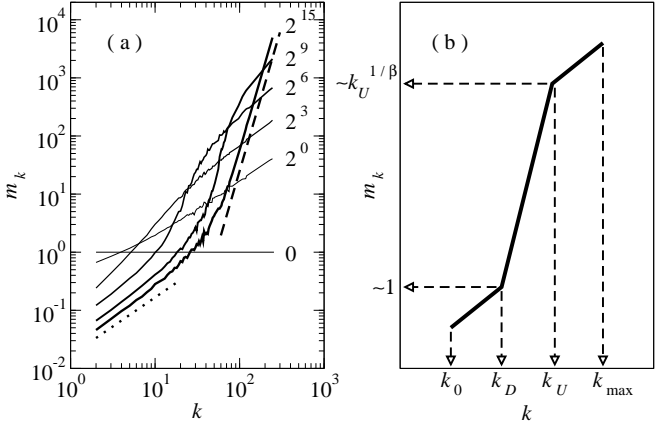}
\caption{ (a) The occupation number distribution $m_k$
at $0 \leq t \leq 2^{15}$. 
The dotted~(dashed) line has the slope 1~($1/\delta=5$). (b)
Schematic plot of $m_k$ vs $k$ in the log-log scale in the transient
time.} \label{fig3}
\end{figure}

In order to get an insight into the nature of the relaxation dynamics, we have
observed the time evolution of the mean occupation number
distribution. The Monte Carlo simulation data are presented in
Fig.~\ref{fig3}~(a). 
They are obtained on the looped SF network of $L=10^4$ nodes
with $\gamma=3$~($\delta_c=1/2$). The particle density is taken to be $\rho=1$.
We take the jumping rate $p(n)=n^\delta$ with $\delta=0.2$ so that
the system is in the condensation phase. The plotted quantity $m_k$
is the ensemble average of the occupation number $n_i$ averaged over
nodes with the degree $k_i=k$.

The data show how the distribution, starting from the uniform 
distribution at $t=0$, approaches the stationary distribution. 
One can see that the initial uniform distribution evolves very quickly~(at
$t=1$) into  a linear distribution $m_k \sim k$.
This is due to the random walk nature of the initial dynamics. 
The initial uniform distribution 
implies that the jumping probability is the same for all particles at $t=0$. 
So they behave as non-interacting random walkers, which has the distribution
$m_k \sim k$. 
After that, the distribution evolves further and approaches
the stationary distribution. The MC data shows that the system has already
reached the stationary distribution given in Eq.~(\ref{m_cond_scale})
at $t=2^{15}$. 

In the transient period, we find that there are three different regimes that
are separated by two time-dependent degree scales denoted by
$k_D$ and $k_U$. 
It turns out that the occupation number distribution behaves in each
regime as
\begin{equation}
m_k \sim \left\{
 \begin{array}{ll}
  k^1 & \mbox{\quad for \quad} k \lesssim k_D \\
  k^{1/\delta} & \mbox{\quad for \quad} k_D \lesssim k \lesssim k_U \\
  k^1 & \mbox{\quad for \quad} k_U \lesssim k \ .
 \end{array}
\right.
\end{equation}
We illustrate the behavior with the schematic plot in Fig.~\ref{fig3}~(b). 
The degree scales grow in time and approach 
$k_D \rightarrow k_c$ with the crossover degree
$k_c$ given in Eq.~(\ref{k_crossover}) and $k_U \rightarrow k_{\max}$
in the stationary state limit.
Furthermore, the behavior of $m_k$ for $k < k_U$ is identical to the
stationary state distribution in Eq.~(\ref{m_cond_scale}) with $k_D$
and $k_U$ playing the role of $k_c$ and $k_{\max}$, respectively.
For nodes with $k>k_U$, $m_k \propto k$. They are still in the
random walk regime.

\begin{figure}[t]
\includegraphics*[width=0.8\columnwidth]{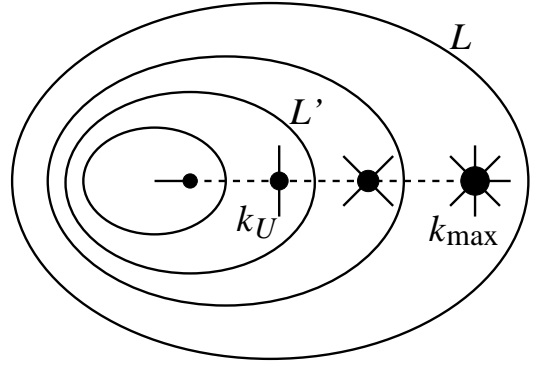}
\caption{Illustration the hierarchical relaxation dynamics. Up to an
intermediate time step $t$, the subnetwork of $L'(t)$ nodes and with
the largest degree $k_U(t) \sim L'(t)^{\beta}$ reaches the
stationary state. The equilibrated area expands as $k_U$ increases
until it reaches $k_{\max}$.} \label{fig4}
\end{figure}

These observations lead us to conjecture that the relaxation
dynamics has a hierarchical nature: A sub-network of 
small-degree nodes  with $k<k_U(t)$ is equilibrated first and 
large-degree nodes become
equilibrated as the degree scale $k_U(t)$ grows in time. In a
transient time $t$, those nodes with $k<k_U(t)$ reach the stationary
state of a smaller network of size $L'(t)\sim k_U(t)^{1/\beta}$ with $\beta
= 1/(\gamma-1)$. The
other degree scale $k_D(t)$ plays the role of the crossover degree
scale in the smaller network, that is,
\begin{equation}\label{kD_kU}
k_D \sim \left\{
\begin{array}{ll}
k_U^{1-\delta/\delta_c} & \mbox{\quad for\quad} \delta<\delta_c \\ [2mm]
(\ln k_U)^{\delta_c} &      \mbox{\quad for\quad} \delta=\delta_c
\end{array}\right.
\end{equation}
following from Eq.~(\ref{k_crossover}). This hierarchical nature
is illustrated in Fig.~\ref{fig4}.

The conjecture predicts that the relaxation dynamics is determined
by the temporal evolution of the degree scale $k_U(t)$. We develop a
scaling theory for $k_U(t)$. Suppose that the system is equilibrated
up to a sub-network of $L'$ nodes with the maximum degree $k_U\sim
L'^\beta$ with $\beta = 1/(\gamma-1)$. 
The node with the maximum degree will be called a
{\em temporary hub}. In order to proceed to the next hierarchy, particles
should be transferred to higher degree nodes. We will assume that
particle transfers from the temporary hub to the next hub dominate
the relaxation time scale, which will be confirmed numerically
shortly. Since the sub-network is in the stationary state, there are
$n \sim L' \sim k_U^{1/\beta}$  particles at the temporary hub.
During the unit time interval, $\Delta n = p(n) \sim L'^\delta$
particles jump out of the temporary hub and perform the random
walk motions. They may wander within the sub-network and return to the
temporary hub, or escape to
the higher degree node in the next hierarchy level. Let
$P_{\text{esc}}(L')$ be the escape probability for the latter
process. Then, we obtain the following scaling relation
\begin{equation}\label{t_formal}
t \sim \frac{1}{P_{\text{esc}}(L')} \times \left(\frac{n}{\Delta n}\right) \sim
\frac{1}{P_{\text{esc}}(L')} \times L'^{1-\delta} \ .
\end{equation}

The escape probability depends on the structure of the underlying networks.
Suppose that the underlying SF network has a tree structure with no loop.
On a tree a path from one node to another is unique. So, particles
can escape from the temporary hub only when they hop along the link
to the right direction toward the next hub among $k_U$ links.
Hence the escape probability is given by
\begin{equation}\label{P_esc_tree}
P_{\text{esc}}(L') \simeq k_U^{-1} \sim L'^{-\beta}
\end{equation}
for a tree network.
On the other hand, in general networks with many loops, there are multiple
pathways between two nodes. Hence, we expect that the escape probability
does not scale algebraically with $L'$ or $k_U$, that is,
\begin{equation}\label{P_esc_nontree}
P_{\text{esc}}(L') \sim L'^0 \sim k_U^0
\end{equation}
for a looped network.
Combining Eqs.~(\ref{t_formal}) (\ref{P_esc_tree}), and
(\ref{P_esc_nontree}), we obtain that
\begin{equation}\label{kUtree}
L' \sim t^{1/( 1 + \beta - \delta )}
\ \ \mbox{or} \ \
k_U \sim t^{\beta / ( 1 + \beta - \delta )}
\end{equation}
for a tree network and
\begin{equation}\label{kUnontree}
L' \sim t^{1/( 1 - \delta )}
\ \ \mbox{or} \ \
k_U \sim t^{\beta / ( 1 - \delta )}
\end{equation}
for a looped network.  The scaling behavior of $k_D$ is obtained from
Eq.~(\ref{kD_kU}).

The system reaches the global stationary state when $L'$ becomes equal to
the network size $L$ or, equivalently, $k_U$ becomes equal to $k_{\max} \sim
L^\beta$. The power-law scaling of $L'$ and $k_U$ implies that
the relaxation time also scales algebraically as
\begin{equation}\label{tau_L_z}
\tau \sim L^z \ .
\end{equation}
The dynamic exponent $z$ depends on the global structure of underlying
networks and is given by
\begin{equation}\label{dynamic_exponent}
z = \left\{
\begin{array}{lll}
 1 + \beta - \delta & , & \mbox{\quad for a tree network} \\
 1 - \delta &,& \mbox{\quad for a looped network.}
\end{array}
\right.
\end{equation}
At the transition $\delta=\delta_c$, there is a logarithmic correction to
the power-law scaling.

We add a remark on the result in Eq.~(\ref{dynamic_exponent}). As explained
in Sec.~\ref{sec:3}, our model at $\delta=0$ has the same stationary state
as the disordered ZRP in periodic lattices with the jumping rate
distribution $P_{\text{jump}}(p') = p'^\nu$ with $\nu = \gamma-2$.
The relaxation dynamics of the ZRP has also been
studied in Refs.~\cite{Krug00,Jain03}. 
It is known to have $z=(\nu+2)/(\nu+1)$
with the biased hopping in one dimension lattices. 
Interestingly, it coincides with 
the result for the ZRP on tree structure SF networks~($\beta=1/(\gamma-1)$).
In the disordered ZRP the relaxation proceeds via a lateral growth of a 
domain within which particles are condensed on to the node with the slowest 
jumping rate. This mechanism is identical to the hierarchical relaxation 
in the SF tree networks. 

\begin{figure}
\includegraphics*[width=\columnwidth]{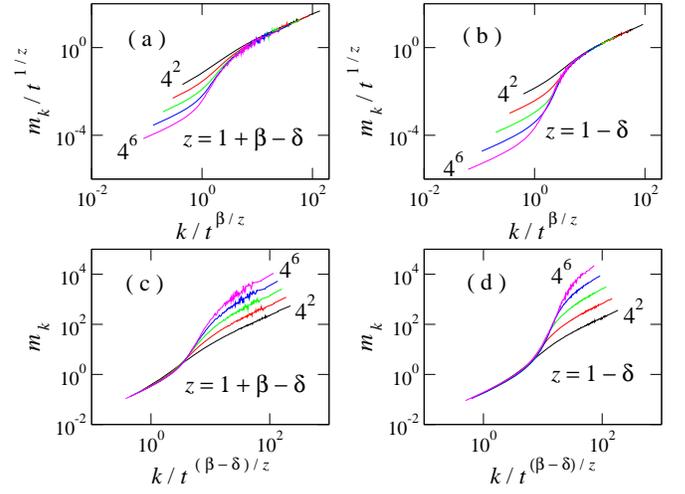}
\caption{(color online) Data collapse analysis of Monte Carlo data of 
the ZRP with $\delta=0.2$ on SF networks of $L=10^5$ nodes and with 
$\gamma=4$. Data in (a) and (c) are from the SF networks with 
the tree structure, while those in (b) and (d) are from the looped 
SF networks. The data are measured in the transient period at
$t=4^2$~(black), $4^3$~(red), $\cdots$, $4^6$~(magenta).}
\label{fig5}
\end{figure}

We verify numerically the analytic results. We have performed the Monte Carlo
simulations on SF networks with various values of $\gamma$ and 
with several values of $\delta$.
In order to study
the structure dependent relaxation dynamics, we have performed the
simulations both on tree networks and looped networks.

According to the scaling argument, the degree scale $k_U(t)$ separating the
second and the third region in Eq.~(\ref{kD_kU}) scales as $k_U(t) \sim
t^{\beta/z}$. And the occupation number at the node with $k=k_U$ 
scales as $m_{k_U} \sim k_U^{1/\beta} \sim t^{1/z}$.
Hence, the existence of $k_U$ can be verified from the
scaling plot of $m_k(t) / t^{1/z}$ against $k / t^{\beta/z}$ at 
transient times $t\ll \tau$.
In Fig.~\ref{fig5}~(a), we present the data for the ZRP with $\delta=0.2$
on a tree-structured SF network of $L=10^5$ nodes and with $\gamma=4$. 
All data at different $t$ collapse very well near $k \sim t^{\beta/z}$ 
with the proposed value of the dynamic exponent 
$z = 1+\beta-\delta$. Figure~\ref{fig5}~(b) shows the data of the ZRP with
the same value of $\delta=0.2$ on a looped SF network with the same value of
$\gamma=4$~($\delta_c = \beta = 1/3$). 
We also observe the perfect data collapse near $k\sim
t^{\beta/z}$ with the proposed value of the dynamic exponent $z=1-\delta$.
These support the structure dependent dynamic scaling of the $k_U$.

Our scaling argument also predicts the existence of the degree scale
$k_D(t) \sim k_U^{1-\delta/\delta_c}$ which plays the role of the crossover
degree $k_c$ for nodes with $k<k_U(t)$. The scaling $k_U(t)\sim t^{\beta/z}$
implies that $k_D \sim t^{(\beta-\delta\beta/\delta_c)/z}$. Note that
$\beta=\delta_c = 1/(\gamma-1)$. Hence, it can be verified with the 
scaling plot of $m_k$ against $k/t^{(\beta-\delta)/z}$. Since the occupation
number at the crossover degree is a constant, the occupation number need
not be rescaled. We present the scaling plots in
Fig.~\ref{fig5}~(c) and (d) for a tree SF network and a looped SF network
with the corresponding dynamic exponent, respectively. Data from different
times $t$ collapse very well near $k\sim t^{(\beta-\delta)/z}$, which
supports the analytic result.
We have also performed the data collapse analysis at several other values of
$\gamma$ and $\delta$, and obtained the consistent result with
the analytic result.

\begin{figure}[t]
\includegraphics*[width=\columnwidth]{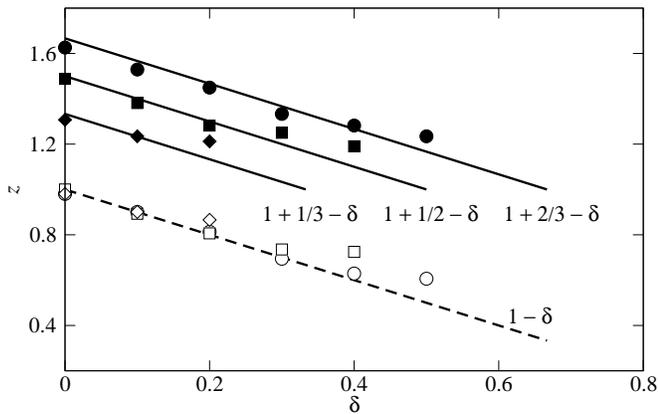}
\caption{Dynamic exponent of tree~(filled symbols) structure SF networks
and looped~(open symbols) structure SF networks. Each symbol stands for 
$\gamma=5/2$~(circle), $\gamma=3$~(square), and $\gamma=4$~(diamond).
The solid and dashed lines represent the analytic
result in Eq.~(\ref{dynamic_exponent}) for tree and looped networks,
respectively.}
\label{fig6}
\end{figure}

In addition to the data collapse analysis, we also measure the dynamic exponent
$z$ through a temporal scaling behavior of $m_{\text{hub}}$. The scaling
picture predicts that nodes with $k<k_U$ are equilibrated while nodes with
$k>k_U$ are in the random walk regime with $m_k \propto k$.
So the occupation number at the hub scales as $m_{\text{hub}} 
\sim k_U^{1/\beta} \times ( k_{\max} / k_U)$. Since $k_U \sim t^{\beta/z}$ 
in the transient times $t\ll \tau$, the occupation number at the hub shows
the temporal scaling 
\begin{equation}
m_{\text{hub}} \sim t^{(1-\beta)/z} \ .
\end{equation}

From the temporal scaling of $m_{\text{hub}}$, we obtained the values of
the  dynamic exponent $z$ at different values of $\gamma$ and $\delta$ on
SF networks of size $L=10^5$ 
of the tree and looped structure. We present the result in
Fig.~\ref{fig6}. All the results are consistent with the analytic results in
Eq.~(\ref{dynamic_exponent}). Approaching $\delta=\delta_c=1/(\gamma-1)$,
the numerical results seem to deviate from the analytic predictions.
It is attributed to strong finite size effects near $\delta=\delta_c$ for we
obtained better results with larger system sizes.

The numerical analysis supports the conjecture of the hierarchical
relaxation and the scaling argument for the structure-dependent
dynamic exponent $z$ in the condensed phase with $\delta \leq \delta_c$.
Our result shows that the dynamical property depends on the global
structure of the underlying networks, although the stationary state property
does not.

\section{Summary}\label{sec:5}
We have studied the stationary and dynamical properties of the ZRP with the
jumping rate $p(n)=n^\delta$ on SF networks with the degree distribution
exponent $\gamma$. The stationary state property is 
determined by the degree distribution of underlying networks.
The model displays the stationary state condensation transition and the
phase diagram is given in Fig.~\ref{fig2}. 
The phase transition is driven by the structural disorder in underlying
scale-free networks. It is distinct from the Bose-Einstein type condensation
transition found in systems with disordered jumping rates, and from the
spontaneous symmetry breaking transition found in systems on periodic
lattices.
In the condensed phase $\delta\le \delta_c = 1 / (\gamma-1)$, 
the mean occupation number distribution follows the scaling form of
Eq.~(\ref{m_cond_scale}). It shows that the whole fraction of particles are
concentrated on higher degree nodes with $k>k_c$ given in
Eq.~(\ref{k_crossover}). On the other hand, the dynamical property depends
also on the global structure of underlying networks. The relaxation dynamics
is found to have the hierarchical nature that it proceeds
from lower degree nodes toward higher degree nodes. It is characterized by
the power-law growth of the degree scale $k_U$. From the scaling argument,
we show that it grows in time with the different exponents depending on
whether underlying networks has a tree or looped structure~(see
Eqs.~(\ref{kUtree}) and (\ref{kUnontree})). Consequently, the dynamic
exponent for the relaxation time $\tau \sim L^z$ also depends on the global
structure of networks~(see Eq.~(\ref{dynamic_exponent})).
Our result shows that the global network structure is the important factor
in understanding the dynamical property of complex networks.

\acknowledgements
This work was supported by Korea Research Foundation Grant
(KRF-2004-041-C00139). The author thanks KIAS for the support during the
visit.


\begin{references}
\bibitem{Evans00} M. R. Evans, Braz. J. Phys. {\bf 30}, 42 (2000).
\bibitem{Evans98} M. R. Evans, Y. Kafri, H. M. Koduvely, and D. Mukamel,
        Phys. Rev. Lett. {\bf 80}, 425 (1998).
\bibitem{Arndt98} P. F. Arndt, T. Heinzel, and V. Rittenberg,
        J. Phys. A {\bf 31}, L45 (1998).
\bibitem{Kafri02} Y. Kafri, E. Levine, D. Mukamel, G.M. Sch\"utz,
        and J. T\"or\"ok, Phys. Rev. Lett. {\bf 89}, 035702 (2002).
\bibitem{Spitzer70} F. Spitzer, Adv. Math. {\bf 5}, 246 (1970).
\bibitem{Grosskinsky03} S. Gro{\ss}kinsky, G. M. Sch\"utz, and H. Spohn
         J. Stat. Phys. {\bf 113}, 389 (2003).
\bibitem{Hanney05} M. R. Evans and T. Hanney, cond-mat/0501338 (2005).
\bibitem{Krug96} J. Krug and P. A. Ferrari, 
         J. Phys. A {\bf 29}, L465 (1996).
\bibitem{Evans96} M. R. Evans, Europhys. Lett. {\bf 36}, 13 (1996).
\bibitem{Krug00} J. Krug, Braz. J. Phys. {\bf 30}, 97 (2000).
\bibitem{Jain03} K. Jain and M. Barma, 
         Phys. Rev. Lett. {\bf 91}, 135701 (2003).
\bibitem{Hwang} K. Huang, {\em Statistical Mechanics}, 2nd ed.
         (Willey, New York, 1987).
\bibitem{Watts98} D. J. Watts and S. H. Strogatz,
        Nature (London) {\bf 393}, 440 (1998).
\bibitem{Albert02} R. Albert and A. -L. Barab\'asi,
        Rev. Mod. Phys. {\bf 74}, 47 (2002).
\bibitem{Dorogov02} S. N. Dorogovtsev and J. F. F. Mendes,
        Adv. Phys. {\bf 51}, 1079 (2002).
\bibitem{Newman03} M. E. J. Newman, SIAM Rev. {\bf 45}, 167 (2003).
\bibitem{Noh04a} J. D. Noh and H. Rieger,
        Phys. Rev. Lett. {\bf 92}, 118701 (2004).
\bibitem{Noh04b} J. D. Noh and H. Rieger,
        Phys. Rev. E {\bf 69}, 036111 (2004).
\bibitem{Barthelemy04} M. Barthelemy, A. Barrat, R. Pastor-Satorras,
        A. Vespignani, Phys. Rev. Lett. {\bf 92}, 178701 (2004).
\bibitem{Gallos04} L. K. Gallos and P. Argyrakis,
        Phys. Rev. Lett. {\bf 92}, 138301 (2004).
\bibitem{Catanzaro05} M. Catanzaro, M. Bogu\~n\'a, and R. Pastor-Satorras,
        Phys. Rev. E {\bf 71}, 056104 (2005).
\bibitem{Sood05} V. Sood and S. Redner,
        Phys. Rev. Lett. {\bf 94}, 178701 (2005).
\bibitem{Noh05} J. D. Noh, G. M. Shim, and H. Lee,
        Phys. Rev. Lett. {\bf 94}, 198701 (2005).
\bibitem{Evans03} M. R. Evans and T. Hanney,
        J. Phys. A {\bf 36}, L441 (2003).
\bibitem{Evans04} M. R. Evans, S. N. Majumdar, and R. K. P. Zia,
         J. Phys. A {\bf 37}, L275 (2004).
\bibitem{comment3} The exact solution is also available when the 
        jumping rate depends on the occupation numbers at both departure 
        and destination nodes in a particular way~\cite{Hanney05}.
\bibitem{Godreche03} C. Godr\`eche, J. Phys. A {\bf 36}, 6313 (2003).
\bibitem{Torok04} J. T\"or\"ok, cond-mat/0407567 (2004).
\bibitem{Grosskinsky04} S. Gro{\ss}kinsky and T. Hanney,
        cond-mat/0412593 (2004).
\bibitem{comment1} In Ref.~\cite{Noh05}, the normalized distribution
        $\omega_i = k_i / (\sum_j k_j )$ was used. However the
        normalization can be ignored.
\bibitem{cutoff} M. Bogu\~n\'a, R. Pastor-Satorras, and A. Vespignani,
        Eur. Phys. J. B {\bf 38}, 205 (2004);
        M. Catanzaro, M. Bogu\~n\'a, and R. Pastor-Satorras,
        Phys. Rev. E {\bf 71}, 027103 (2005).
\bibitem{comment2} In Ref.~\cite{Noh05}, only the second term was
considered. However, this does not change any result.
\bibitem{YKim05} Y. Kim, unpublished (2005).
\bibitem{DMS} S.N. Dorogovtsev, J.F.F. Mendes, and A.N. Samukhin,
        Phys. Rev. Lett. {\bf 85}, 4633 (2000).
\bibitem{BAnet} A.-L. Barab\'asi and R. Albert,
                Science {\bf 286}, 509 (1999);
        A.-L. Barab\'asi, R. Albert, and H. Jeong,
                Physica A {\bf 272}, 173 (1999).

\end{references}
\end{document}